\documentclass[preprint]{revtex4}  
\usepackage{amsfonts}
\usepackage{amsmath}
\usepackage{cancel}
\usepackage{graphicx}
\usepackage{mathcomp}
\usepackage{color}
\usepackage{ulem}
\usepackage{paralist}
\usepackage{gensymb}
\usepackage{placeins}
\usepackage{hyperref}

\newif\iffigures
\figurestrue

\def\undertilde#1{\mathord{\vtop{\ialign{##\crcr
$\hfil\displaystyle{#1}\hfil$\crcr\noalign{\kern1.5pt\nointerlineskip}
$\hfil\widetilde{}\hfil$\crcr\noalign{\kern1.5pt}}}}}

\usepackage[deletedmarkup=xout]{changes}
\definechangesauthor[color=green]{RPB}
\definechangesauthor[color=cyan]{SW}
\definechangesauthor[color=orange]{JL}
\definechangesauthor[color=blue]{TM}
\definechangesauthor[color=magenta]{MK}

\begin{document}
\title{Validation of the smooth step model \\ by particle-in-cell/Monte Carlo collisions simulations}

\author{ Maximilian Klich, Jan Löwer, Sebastian Wilczek, Thomas Mussenbrock, Ralf Peter Brinkmann}

\affiliation{Department of Electrical Engineering and Information Technology \\ 
						 Ruhr University Bochum, D-44780 Bochum, Germany}

\date{\today}

\begin{abstract}

Bounded plasmas are characterized by a rapid but smooth transition from quasi-neutrality in the volume to
electron depletion close to the electrodes and chamber walls. The thin non-neutral region,  \linebreak
the boundary sheath,
comprises only a small fraction of the discharge domain but controls much of its macroscopic behavior.
Insights into the properties of the sheath and its relation to the plasma are of high practical and theoretical interest.
The recently proposed smooth step model provides a closed analytical expression for the electric field in a 
planar, 
radio-frequency modulated  sheath. 
\linebreak It represents (i) the space charge field in the depletion zone, 
(ii) the generalized Ohmic and ambi\-polar field in the quasi-neutral zone, 
and (iii) a smooth interpolation for the transition in between. 
This investigation compares the smooth step model with the predictions of a more fundamental \linebreak
 particle-in-cell/Monte Carlo 
collisions simulation and finds good quantitative 
agreement when the assumed length and time scale requirements are met.
A second simulation case illustrates that the model 
 remains applicable even when the assumptions are only
marginally fulfilled.

\end{abstract}

\maketitle

\pagebreak


\section{Introduction}

In radio-frequency (RF) discharges, the electron-depleted sheath occupies only a fraction of \linebreak 
the volume but governs many 
of the phenomena. 
Its electric field exceeds that of the plasma by more than three orders of magnitude and
plays an important role in the processes of particle acceleration and power absorption.
The relation between the sheath charge $Q$ and the sheath 
voltage $V$ controls the 
discharge impedance.
 A solid understanding
of the physics  of the sheath is thus of technological value. Much research bears witness to this
\cite{Child1911,Langmuir1913,MottGurney1940,Warren1955,Bohm1949}.

Theoretical models of the sheath dynamics can be formulated in varying levels of complexity, \linebreak
ranging from lumped element diode models to first-principles based numerical simulations \linebreak
which often employ the particle-in-cell/Monte Carlo collisions (PIC/MCC) algorithm \cite{MetzeErnieOskam1986,ShihabZieglerBrinkmann2012}.
\linebreak
This study addresses the middle ground. Semi-analytic sheath models adopt mathematical 
simplifications but keep the essential physics. Such models were first analyzed in  
the 1980s \linebreak by 
Godyak et al.~\cite{GodyakGhanna1979,Godyak1986,GodyakSternberg1990} and Lieberman \cite{Lieberman1988,Lieberman1989}. 
Using the fluid approach, they assumed a two-species plasma with electrons and singly charged ions in the RF regime $\omega_\mathrm{pi}\!\ll\!\omega_\mathrm{RF}\!\ll\! \omega_\mathrm{pe}$.  
\linebreak
(Here, $\omega_\mathrm{RF}$ is the RF frequency, while $\omega_\mathrm{pi}$ and $\omega_\mathrm{pe}$ are the 
plasma frequencies of the ions and the electrons, respectively.) 
Many studies adopted these assumptions  \cite{Riemann1989,Czarnetzki2013}.

\begin{figure}[t!]
\centering
\includegraphics[width=\textwidth]{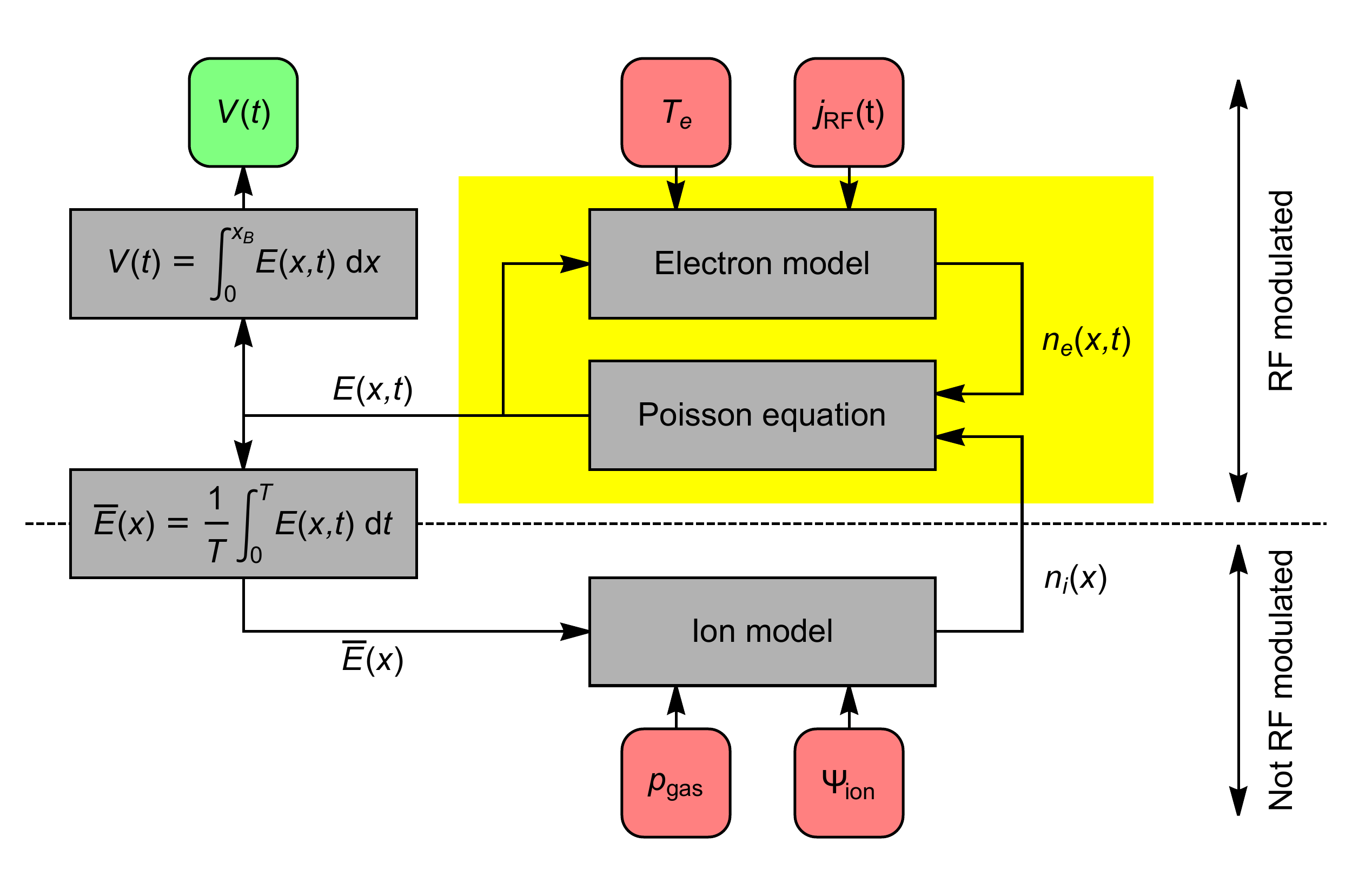}
\caption{Structure of an RF boundary sheath model. The RF modulated sector contains the electron model and Poisson's equation which form a system of partial differential equations in $x$ and $t$.\linebreak 
The unmodulated sector contains the ion model, a system of ordinary differential equations~in~$x$. \linebreak 
The two sectors are coupled via an integral relation which maps the phase-resolved field $E(x,t)$ 
onto the phase-averaged field $\bar{E}(x)$.
Semi-analytic RF boundary sheath models replace the electron and field equations by their approximate solution, i.e., by an effective field model
(yellow sector).\linebreak In the tradition started by Godyak and Lieberman,
this effective field model is the HSM
\cite{GodyakGhanna1979,Godyak1986,GodyakSternberg1990,Lieberman1988,Lieberman1989}. 
It~is~advantageous to substitute it by the SSM \cite{Brinkmann2007,Brinkmann2015,Brinkmann2016}.}
	\label{FluidSheathModel}
\end{figure}

The general structure of such sheath models is shown in Fig.~\ref{FluidSheathModel}. As the RF regime is adopted, \linebreak 
the model can be separated in two sectors, one that is phase-resolved and one that is not. 
\linebreak
The phase-resolved sector contains the electron model and Poisson's equation which solve for the electron density
$n_\mathrm{e}(x,t)$ and the electric field $E(x,t)$. The phase-averaged field $\bar{E}(x)$  \linebreak
reports to the non-phase-resolved sector, where the ion model solves for the ion density $n_\mathrm{i}(x)$.\linebreak
This quantity, in turn, is communicated to the phase-resolved sector. Various quantities must \linebreak
 be provided as external input,
most notably the RF modulation, the flux density of the ions, 
and the composition, pressure, and temperature of the neutral gas.

RF sheath models are mathematically complicated, and drastic simplifications are called for.
Godyak and Lieberman found such a simplification. Noting that 
the passage from electron
depletion to quasineutrality in a plasma-sheath transition
is steep, they approximated the electron density by an
infinitely sharp front (hard step) located at the electron edge $s(t)$.
This allowed integrating Poisson's equation directly. 
The resulting hard step model (HSM) provides a closed formula for the  electric field $E(x,t)$ in terms of 
the sheath charge $Q(t)$ and the ion density $n_\mathrm{i}(x)$ and simplifies the set-up of sheath models considerably, see Fig.~\ref{FluidSheathModel}.

Since then, the HSM has become very popular, and was used in many 
investigations \cite{Riemann1989,Czarnetzki2013}. \linebreak
However, it is a rather drastic approximation and has, consequently, some severe drawbacks. \linebreak
Implicitly, the HSM sets the electron temperature equal to zero, which renders it impossible \linebreak to describe 
the ambipolar field and the floating voltage. Furthermore, it assumes that the electrons can  follow the electric field instantaneously;
this excludes dynamic phenomena like the action of inertia, 
the emergence of an Ohmic field, and the 
effect of field reversal~\cite{SchulzeDonkoHeilLuggenhoelscherMussenbrockBrinkmannCzarnetzki2008}.

\pagebreak

In a recent sequence of studies, one of the current authors proposed an improved formula \linebreak
for the electric field in a sheath-plasma transition, the smooth step model (SSM) 
\cite{Brinkmann2007,Brinkmann2015}. \linebreak
This advanced field model accounts for thermal and dynamic effects in terms of a higher-order perturbation analysis. 
 It represents (i) the space charge field in the depletion zone, \linebreak
(ii) the generalized Ohmic and ambi\-polar field in the quasi-neutral zone, 
and (iii) a smooth interpolation for the transition in between.
The SSM  was applied, for example, to set up a consistent theory of electron heating \cite{Brinkmann2016}.

Nonetheless, the SSM is only an approximation. How does it compare to the outcome of a more fundamental
(and numerically much more resource-consuming) PIC/MCC-simulation? 
\linebreak
This is the main question of this study,
and we will proceed as follows: In the next section, \linebreak
 we sketch the   
basis of all plasma modeling, kinetic theory, which couples a set of Boltzmann equations
for the particle distribution functions to Poisson's equation for the electric field. 
The PIC/MMC algorithm to solve the kinetic model in a stochastic sense is briefly reviewed.
\linebreak
Section III provides a fluid dynamic parent model as the basis of both the HSM and the SSM, \linebreak
and section IV sketches the corresponding derivations. Section V introduces two PIC/MCC simulation case of a planar capacitively coupled discharge, a single frequency case (1f) and a double frequency case (2f).
The cases are evaluated with focus on the sheath dynamics.
The electrical field and the sheath voltages
predicted by the two models are compared with the results of the PIC/MCC studies.  
The manuscript concludes with a brief conclusion and an outlook in section IV. 

\pagebreak

\section{A kinetic model of a planar discharge}

Kinetic theory gives the most fundamental model of low temperature gas discharges 
\cite{LiebermanLichtenberg2005,MakabePetrovic2015}. \linebreak
We assume a plane-parallel geometry with Cartesian coordinates. 
The $x$-axis points from the electrode at the position $x=0$ to the bulk. 
The opposite electrode is located at $x=L$. \linebreak
Translational invariance in $y$ and $z$ as well as axisymmetry around the $x$-axis are assumed.
To~avoid having to define problematic boundary conditions at the sheath-plasma transition, we treat the whole discharge 
domain $[0,L]$ and select as the sheath edge $x_\mathrm{B}$ the first point that is always quasineutral in the RF period. 
The ion density at the point $x_\mathrm{B}$ will be called $n_\mathrm{B}$. \linebreak
(The slight arbitrariness in these definitions is harmless, as the gradients at $x_\mathrm{B}$ are small.) \linebreak
The discharge is periodic in the RF period $T_\mathrm{RF} = 2\pi \omega_\mathrm{RF}$. 
Kinetic equations are formulated for the distribution functions $f_s(x,v_\parallel,v_\perp,t)$,
$s= 1\ldots N_\mathrm{s}$, where
$s=1 \equiv \mathrm{e}$ refers to 
electrons and $s\ge 2$ to ions.
We use cylindrical velocity coordinates,  $v_\parallel \equiv v_x$ and $v_\perp = \sqrt{v_y^2+v_z^2}$.
Moreover, we abbreviate $E \equiv E_x(x,t)$.
The term on the right is the collision term to represent elastic and inelastic collisional interaction and chemistry:
\begin{align}
	  \frac{\partial f_s}{\partial t} + v_\parallel  \frac{\partial f_s}{\partial x} +
		\frac{q_s}{m_s} E  \frac{\partial f_s}{\partial v_\parallel} =  \frac{\partial f_s}{\partial t}\Bigl\vert_\mathrm{c}.
		\label{KineticEquation}
\end{align}
At $x=0$ and $x=L$, boundary conditions are posed. 
We neglect the emission of secondary particles due to electron or ion impact; 
instead we assume ideal absorption:
\begin{align}
    f_s(x,v_\parallel,v_\perp,t)\vert_{x=0} = 0\;\; \mathrm{for} \; v_\parallel>0, \\[0.25ex]
		f_s(x,v_\parallel,v_\perp,t)\vert_{x=L} = 0 \;\; \mathrm{for} \; v_\parallel<0.
\end{align}
We adopt the electrostatic approximation; the electrical field can be expressed by a potential. 
Poisson's equation applies; the term on the right is the charge density:
\begin{align}
	 \varepsilon_0 \frac{\partial E}{\partial x} = - \varepsilon_0 \frac{\partial^2\Phi}{\partial x^2} = \rho(x,t) = 
	  \sum_{s=1}^{N_\mathrm{s}} q_s \int_{-\infty}^\infty\int_0^\infty 
		f_s(x,v_\parallel,v_\perp,t) \, 2\pi\, v_\perp \,  \mathrm{d}v_\perp\, \mathrm{d}v_\parallel. 
\end{align}
The boundary conditions for the potential are as follows, where the RF voltage $V_\mathrm{RF}(t)$ is periodic and phase-average free but not 
necessarily harmonic.
We assume floating conditions, i.e., demand that also the discharge current $J(t)$
is phase-average free. 
The voltage $V_\mathrm{b}$ is the constant self-bias voltage:
\begin{align}
    &\Phi(0,t)\vert_{x=0} = 0, \\[0.25ex]
		&\Phi(L,t)\vert_{x=L} = V_\mathrm{b} + V_\mathrm{RF}(t). 	\label{PotentialBC}
\end{align}

The particle-in-cell/Monte Carlo collisions algorithm provides self-consistent solutions of the  
kinetic model \eqref{KineticEquation} - \eqref{PotentialBC} in a stochastic sense 
\cite{Birdsall1991, VenderBoswell1992, Verboncoeur2005,TskhakayaMatyashSchneiderTaccogna2007}. 
It resolves nonlocal and nonlinear effects that are important for low-pressure plasmas
and has become a commonly used tool  
\cite{WilczekTrieschmannEreminBrinkmannSchulzeSchuengelDerzsiKorolovHartmannDonkoMussenbrock2016, SchulzeDonkoLafleurWilczekBrinkmann2018,
WilczekSchulzeBrinkmannDonkoTrieschmannMussenbrock2020, Horvath, Derzsi,  PICbenchmark, DonkoPIC}.
 \linebreak
The algorithm combines a particle-based scheme for the kinetic equations with a grid-based 
representation of the electric field. 
In the particle step, an ensemble of superparticles,
\linebreak 
following the Newton equations of motion, is propagated forward by a small time step $\Delta t$.
\linebreak
This allows to track individual particles and avoids the problem of numerical diffusion.
Mathematically, the characteristics of the kinetic equation are followed.
Each superparticle represents $w_s$ physical particles.
Masses are rescaled as $m_s \to w_s m_s$, charges as $q_s \to w_s q_s$,\linebreak
densities as $n_s \to n_s/w_s$, 
fluxes~as~$\psi_s \to \psi_s/w_s$, 
and particle energies as $T_s \to w_s T_s$,
\linebreak
keeping the charge-to-mass ratio $q_s/m_s$, 
the Debye length $\lambda_\mathrm{D}$,
 and the plasma frequency $\omega_\mathrm{pe}$.\linebreak
Individual inter\-actions 
of particles with the neutral background and the reactor walls
are accounted for by  Monte Carlo collisions using the null collision method \cite{Skullerud1968}. 
For the field step, \linebreak
the charges are mapped to a grid of cell size $\Delta x$.
The calculated fields become interpolated at the particles' position and the cycle starts again, repeated until a periodic state is assumed.
\linebreak
To ensure stability and accuracy, the  conditions $\Delta t \!\le\! 0.2/\omega_\mathrm{pe}$ and $\Delta x\! \le \! 0.5 \lambda_\mathrm{D}$
must hold \cite{PICbenchmark}.
\linebreak
Various diagnostics then calculate 
properties of interest, averaged over a high number of subsequent PIC/MCC cycles for better statistics. 
Remaining small-scale fluctuations are cleared with the help of a discrete diffusion algorithm. 

Of particular importance for the interpretation of a discharge simulation are the velocity moments and the 
corresponding balance equations. All~moments are functions of $x$ and $t$; we list them up to the second order.
The moments of order zero are the densities:
\begin{align}
	 n_s =  \int_{-\infty}^\infty\int_0^\infty f_s \, 2\pi\, v_\perp \, \mathrm{d}v_\perp\, \mathrm{d}v_\parallel.
\end{align}
The first order moments appear as the flux densities $ \psi_s$ or mean velocities $u_s$:  
\begin{align}
	 \psi_s = n_s u_s =\int_{-\infty}^\infty\int_0^\infty  v_\parallel\,f_s \, 2\pi\, v_\perp \, \mathrm{d}v_\perp\, \mathrm{d}v_\parallel.
\end{align}
The moments of order two are the parallel and perpendicular pressures $p_{\parallel s}$ and $p_{\perp s}$ or the
parallel and perpendicular temperatures $T_{\parallel s}$ and $T_{\perp s}$, respectively:
\begin{align}
	& p_{\parallel s} =  n_s T_{\parallel s}  =
	\int_{-\infty}^\infty\int_0^\infty  m_s
	          \left( v_\parallel - u_s\right)^2  f_s \, 2\pi\, v_\perp \, \mathrm{d}v_\perp\, \mathrm{d}v_\parallel, \\[0.25ex]
	&	 p_{\perp s} =  n_s T_{\perp s}  =
	\int_{-\infty}^\infty\int_0^\infty \frac{1}{2} m_s
	          v_\perp^2  f_s \, 2\pi\, v_\perp \, \mathrm{d}v_\perp\, \mathrm{d}v_\parallel.				
\end{align}

Of the balance equations, the first is the particle balance or equation of continuity:
\begin{align}
	 \frac{\partial  n_s}{\partial t} + \frac{\partial \psi_s}{\partial x} =  S_s.
\end{align}
The collisional source terms on the right are defined as
\begin{align}
&	 S_s =  \int_{-\infty}^\infty\int_0^\infty \frac{\partial f_s}{\partial t}\Bigl\vert_\mathrm{c} \, 2\pi\, v_\perp \, \mathrm{d}v_\perp\, \mathrm{d}v_\parallel.
\end{align}
The law of charge conservation is described by
\begin{align}
     \sum_{s=1}^{N_s} q_s  S_s  = 0.
     \label{ChargeConservation}
\end{align}
The momentum balances or equations of motion are
\begin{align}
	 \frac{\partial }{\partial t}\left( m_s \psi_s \right)+ 
	\frac{\partial}{\partial x}\left( m_s \frac{\psi_s \psi_s}{n_s} + n_s T_{\parallel s}\right) =
	  q_s n_s E + \Pi_s.
\end{align}
The collision induced changes in the momentum density are: 
\begin{align}
	 \Pi_s =\int_{-\infty}^\infty\int_0^\infty
	m _s v_\parallel  \frac{\partial f_s}{\partial t}\Bigl\vert_\mathrm{c} \, 2\pi\, v_\perp \, \mathrm{d}v_\perp\, \mathrm{d}v_\parallel.
\end{align}
In the PIC/MCC simulation, the velocity moments are calculated by summing the respective quantities over all particles in a particular cell.
\pagebreak

\section{A fluid parent model}

As stated in the introduction, the hard step model (HSM) and the smooth step model (SSM) are both based on a fluid approach. 
This term describes a family of less fundamental plasma theories that are based on the first velocity moments of the distribution functions and the corresponding balance equations.
The individual fluid theories differ in how many moments are included and how exactly the system of equations is truncated. In this section we describe  the leanest fluid model that can serve as a common parent of both the HSM and the SSM. \linebreak
The first step of its derivation is to lump the ion species into a compound fluid by defining the density $n_\mathrm{i}$ and 
the flux density $\psi_\mathrm{i}$ as
\begin{align}
	 n_\mathrm{i} &= \frac{1}{e} \sum_{s=2}^{N_\mathrm{s}} q_s n_s,
	\\
   \psi_\mathrm{i} &= \frac{1}{e} \sum_{s=2}^{N_\mathrm{s}} q_s \psi_s. 
\end{align}
For the generation rate, the charge conservation law \eqref{ChargeConservation} allows to write
\begin{align}
    S_\mathrm{i} &= \frac{1}{e} \sum_{s=2}^{N_\mathrm{s}} q_s  S_s = S_\mathrm{e} \equiv S.
\end{align}
Then, the electron and ion equations of continuity are
\begin{align}
	\frac{\partial n_\mathrm{e}} {\partial t}+ \frac{\partial \psi_\mathrm{e}} {\partial x} = S,\label{FullElectronContinuity}\\[0.5ex]
	\frac{\partial n_\mathrm{i}} {\partial t}+ \frac{\partial \psi_\mathrm{i}} {\partial x} = S,
\end{align}
and Poisson's equation reads
\begin{align}
	 \varepsilon_0 \frac{\partial E}{\partial x} = 
	  e(n_\mathrm{i}-n_\mathrm{e}).
\end{align}
As a result, 
the total current $J(t)$, the sum of the particle and the displacement currents,\linebreak is solely a function of time, not position:
\begin{align}
	 J(t) =\varepsilon_0 \frac{\partial E}{\partial t} - e \psi_\mathrm{e} + e \psi_\mathrm{i}. 
\label{CurrentDensity}
\end{align}
Furthermore, the sheath charge 
$Q(t)$, defined as the integral of the charge density 
from the electrode to the bulk point $x_\mathrm{B}$, can be expressed as the
difference of the electric fields:
\begin{align}
	  Q(t) = \int_{0}^{x_\mathrm{B}}  e (n_\mathrm{i}-n_\mathrm{e}) \, \mathrm{d}x  = \varepsilon_0 \bigl(E(x_\mathrm{B},t) - E(0,t) \bigr).
	   \label{ChargeBalance}
\end{align}

Because of their large mass, the ions solely see the phase-averaged field. Their density is a time-independent function which,
as~indicated in Fig.~\ref{FluidSheathModel},
can be considered an input:
\begin{align}
    n_\mathrm{i}(x)
      = \text{given positive and monotonous function}.
\end{align}
It is useful to introduce the so-called charge coordinate $q(x)$, 
\begin{align}
	 q(x) &= \int_0^{x} e n_\mathrm{i}(x^\prime)\, \mathrm{d}x^\prime. \label{ChargeCoordinate}
\end{align}
The light electrons are strongly modulated. 
The ionization term $S$ in \eqref{FullElectronContinuity}
is thus negligible compared to the other two terms.
Furthermore, only the fluctuating (phase-average free) \linebreak 
part of the electron flux enters the equation of continuity:
\begin{align}
    \frac{\partial n_\mathrm{e}} {\partial t}+ \frac{\partial \tilde{\psi}_\mathrm{e}} {\partial x} = 0.\label{ReducedElectronContinuity}
\end{align}
We simplify the electron momentum balance by interpreting the interaction of electrons with neutrals as friction and neglecting the non-fluctuating part of the electron flux: 
\begin{align}
	 \frac{\partial }{\partial t}\left( m_\mathrm{e} \tilde{\psi}_\mathrm{e} \right)+ 
	\frac{\partial}{\partial x}\left( m_\mathrm{e}  \frac{\tilde{\psi}_\mathrm{e}^2}{n_\mathrm{e}} + n_\mathrm{e} T_{\mathrm{e}}\right) =
	  -e  n_\mathrm{e} E - \nu_\mathrm{e} m_\mathrm{e} \tilde{\psi}_\mathrm{e}. 
		\label{ElectronMotionFluid}
\end{align}
The electron friction constant is treated as an input:
\begin{align}
    \nu_\mathrm{e} = \text{given positive constant.} 
\end{align}
Also the electron temperature is treated as an input. Traditionally assumed to be a constant, it may more generally be a function provided by other parts of the model:
\begin{align}
    T_\mathrm{e}(x,t) = \text{given positive, RF-periodic function.}
\end{align}
The floating condition implies that the phase-averaged flux of the electrons $\bar{\psi}_\mathrm{e}$ and of the ions $\bar{\psi}_\mathrm{i}$ must be equal at the electrode.
As $\bar{\psi}_\mathrm{e}$ 
has already been dropped from the model, 
the Hertz-Langmuir relation \cite{Riemann1989} 
is employed: 
\begin{align}
 |\bar{\psi}_\mathrm{e}|\big|_{x=0} = \sqrt{\frac{T_\mathrm{e}}{2\pi m_\mathrm{e}}} \, \bar{n}_\mathrm{e}\big|_{x=0} \stackrel{\displaystyle !}{=} 
 |\psi_\mathrm{i}|\big|_{x=0} = \text{given constant}.
 \label{PhaseAverageCurrent}
\end{align}
Taking the time derivative of \eqref{ChargeBalance},
invoking \eqref{CurrentDensity} and \eqref{PhaseAverageCurrent}, and neglecting the
fluctuating components of the electron flux at  $x=0$ and the displacement current at $x_\mathrm{B}$,
we get
\begin{align}
	 \frac{\mathrm{d}Q}{\mathrm{d}t}  = -J(t).
		\label{QBalance}
\end{align} 
The RF current $J(t)$ is treated as an input:
\begin{align}
    J(t) = \text{given RF periodic, phase-average free function.}
\end{align}

\pagebreak

\section{Hard step model and smooth step model}

As discussed, semi-analytical sheath models require a closed formula 
to express the electric field in the sheath-to-plasma transition in terms of the ion density and the sheath charge. \linebreak
The HSM and the SSM provide such closed formulas. 
They start from the same assumptions regarding the spatial and temporal scales,  
here formulated in the terminology of \cite{Brinkmann2015}:
\begin{itemize}
	\item The transition from electron depletion to quasi-neutrality occurs within a thin zone.
        The governing scale of this transition is the Debye length $\lambda_\mathrm{D}$, while the sheath
         extension \linebreak scales with the gradient lengths $l$ of the ion density and the electron temperature $T_\mathrm{e}$.
		The corresponding smallness parameter is termed $\epsilon$: 				
\begin{align}
	  \epsilon= \frac{\lambda_\mathrm{D}}{l} = 
		\sqrt{\frac{\varepsilon_0 T_\mathrm{e}}{e^2 n_\mathrm{i}}}\frac{1}{n_\mathrm{i}}\frac{\partial n_\mathrm{i}}{\partial x}
			\sim \sqrt{\frac{\varepsilon_0 T_\mathrm{e}}{e^2 n_\mathrm{i}}}\frac{1}{T_\mathrm{e}}\frac{\partial T_\mathrm{e}}{\partial x}
			\ll  1.
\end{align}
  \item The RF modulation frequency $\omega_\mathrm{RF}$ is small compared to the plasma frequency $\omega_\mathrm{pe}$. \linebreak
	      Thus, the electrons behave quasi-statically. The electron collision frequency $\nu_\mathrm{e}$ 
				and the modulation rate of the electron temperature $T_\mathrm{e}$ are assumed to be 
				comparable to $\omega_\mathrm{RF}$. \linebreak
				The corresponding smallness parameter is termed $\eta$: 
        \begin{align}
	         \eta = \frac{\omega_\mathrm{RF}}{\omega_\mathrm{pe}} = 
					\frac{\omega_\mathrm{RF}}{\sqrt{e^2 n_\mathrm{i}/\varepsilon_0 m_\mathrm{e}}} \ll 1.
        \end{align}
	
\end{itemize}

The HSM implements these assumptions in a drastic way; the smallness parameters $\epsilon$ and $\eta$\linebreak
are both set equal to zero. Equivalently, one can simply set the electron temperature $T_\mathrm{e}$ and the 
electron mass $m_\mathrm{e}$ equal to zero.
The equation of motion \eqref{ElectronMotionFluid} then reduces to $n_\mathrm{e} E = 0$.
To solve this equation, the ansatz of a hard step in the electron density is made:
\begin{align}
	n_\mathrm{e,\,HSM}(x,t)=\left\{\begin{array}{cll}
0 & : & x<s(t) \\[0.5ex]
n_\mathrm{i}(x) & : & x>s(t)
\end{array}\right. .
\end{align}
Equation \eqref{QBalance} is time-integrated under the condition that the minimal value of $Q(t)$ is zero; \linebreak
this corresponds to the asymptotic form of the Hertz-Langmuir relation \eqref{PhaseAverageCurrent} 
under the assumptions of the HSM.
The location of the hard step $s(t)$, with a minimum that is also zero,\linebreak is determined by the sheath charge $Q(t)$ via
\begin{align}
	  Q(t) =  \int_0^s e n_\mathrm{i}(x^\prime) \,\mathrm{d}x^\prime, \label{SheathChargeHSM}
\end{align}

Integrating Poisson's equation, one can then express the electric field of the HSM as follows. \linebreak
In the unipolar region, the field reflects the charge between $s(t)$ and $x$. 
In the ambipolar zone, \linebreak it is, due to $T_\mathrm{e} \to 0$ and  $m_\mathrm{e} \to 0$, identical to zero.  
The space charge  field is accounted for,\linebreak but thermal and dynamical effects are completely neglected: 
\begin{align}
E_\mathrm{HSM}(x,t) = \frac{1}{\varepsilon_0}\min\left(q(x)-Q(t),0\right)=\left\{\begin{array}{ll}
	\displaystyle - \frac{e}{\varepsilon_{0}} \int_{x}^{s} n_\mathrm{i}\left(x^{\prime}\right) d x^{\prime} & : x<s(t) \\[0.5ex]
	 0 & : x>s(t)
	\end{array}\right. .  \label{EHSM}
\end{align}

The SSM treats the parent model equations 
more cautiously.
An asymptotic series in $\epsilon$ and $\eta$ is formulated and truncated after the quadratic order
(\cite{Brinkmann2015} offers a more detailed description). \linebreak
In physical units, the electrical field provided by the SSM reads
\begin{align}
E_{\mathrm{SSM}}(x, t)=&-\sqrt{\frac{n_{\mathrm{i}} T_{\mathrm{e}}}{\varepsilon_{0}}} \Xi_{\mathrm{S}}\!\left(\frac{q(x)-Q(t)}{\sqrt{\varepsilon_{0} n_{\mathrm{i}} T_{\mathrm{e}}}}\right) 
-\frac{1}{e n_{\mathrm{i}}} \frac{\partial}{\partial x}\left(n_{\mathrm{i}} T_{\mathrm{e}}\right) \Xi_{\mathrm{A}}\!\left(\frac{q(x)-Q(t)}{\sqrt{\varepsilon_{0} n_{\mathrm{i}} T_{\mathrm{e}}}}\right) \label{ESSM} \\[1.0ex] \nonumber
&+\frac{m_{\mathrm{e}}}{e^{2} n_{\mathrm{i}}}\left(\frac{\partial J}{\partial t}+\nu_{\mathrm{e}} J-\frac{1}{e} \frac{\partial}{\partial x}\left(\frac{J^{2}}{n_{\mathrm{i}}}\right)\right) \Xi_{\Omega}\!\left(\frac{q(x)-Q(t)}{\sqrt{\varepsilon_{0} n_{\mathrm{i}} T_{\mathrm{e}}}}\right).
\end{align}
The sheath charge $Q(t)$ is obtained by solving  
\eqref{QBalance} under the Hertz-Langmuir relation \eqref{PhaseAverageCurrent}.\linebreak
The special functions $\Xi_\mathrm{S}$,  $\Xi_\mathrm{A}$, and $\Xi_{\Omega}$ (displayed in \cite{Brinkmann2015})
are smooth -- in fact analytical -- 
functions of their argument $\xi$.
By defining the formal electron edge $s(t)$ via 
equation \eqref{SheathChargeHSM}, one can interpret
$\xi$ as a scaled distance of the position $x$ 
to that point \cite{Brinkmann2007}:
\begin{align}
	 \xi = \frac{q(x)-Q(t)}{\sqrt{\epsilon_{0} n_{\mathrm{i}} T_{\mathrm{e}}}} \approx \frac{x-s(t)}{\lambda_\mathrm{D}}.
\end{align}
The SSM is a representation of the electric field which accounts for
thermal and dynamic effects in leading order perturbation theory. It therefore eclipses the HSM:
\begin{itemize}
	\item In the unipolar region left of the transition, $q(x) \ll Q(t)$. The function $\Xi_\mathrm{S}$ becomes
	      equal to its negative argument,  while $\Xi_\mathrm{A}$ and $\Xi_{\Omega}$ vanish. 
				The resulting form describes, \linebreak asymptotically exact, the space charge field and coincides with the HSM:
		\begin{align}
            E_{\mathrm{S}}(x, t)= \frac{q(x)-Q(t)}{\varepsilon_0}. \label{E_S}
        \end{align}		

	\item In the ambipolar region right of the transition, $q(x) \gg Q(t)$. The function $\Xi_\mathrm{S}$ vanishes,  
	     while $\Xi_\mathrm{A}$ and $\Xi_{\Omega}$ become unity. The field reduces to what is known as the 
			 generalized Ohmic field including the ambipolar field \cite{GeneralizedOhmsLaw}:
\begin{align}
E_{\mathrm{A\Omega}}(x, t)=
-\frac{1}{e n_{\mathrm{i}}} \frac{\partial}{\partial x}\left(n_{\mathrm{i}} T_{\mathrm{e}}\right) 
+\frac{m_{\mathrm{e}}}{e^{2} n_{\mathrm{i}}}\left(\frac{\partial J}{\partial t}+\nu_{\mathrm{e}} J-\frac{1}{e} \frac{\partial}{\partial x}\left(\frac{J^{2}}{n_{\mathrm{i}}}\right)\right) .
\end{align}
\item In the transition regime, the formula provides a smooth transition.
\end{itemize}

\pagebreak

\section{Results: Comparison of the different models}

For this investigation, we ran a single-frequency (1f) and a double-frequency (2f) argon case,
using the benchmarked electrostatic code \textit{yapic1D} \cite{PICbenchmark} with Phelps cross sections  
\cite{Phelps, lxcat1, lxcat2, lxcat3}.
Case (1f) was intensively studied as the ``base case'' of a tutorial on plasma simulation \cite{WilczekSchulzeBrinkmannDonkoTrieschmannMussenbrock2020}, 
case (2f) has a different excitation and a lower pressure.
All sheath quantities refer to the sheath at the left electrode $x=0$.

\vspace{-15pt}
\subsection{Single frequency case (1f)}
\vspace{-10pt}

\begin{figure}[h!]
    \centering
    \includegraphics[width=1.0\textwidth]{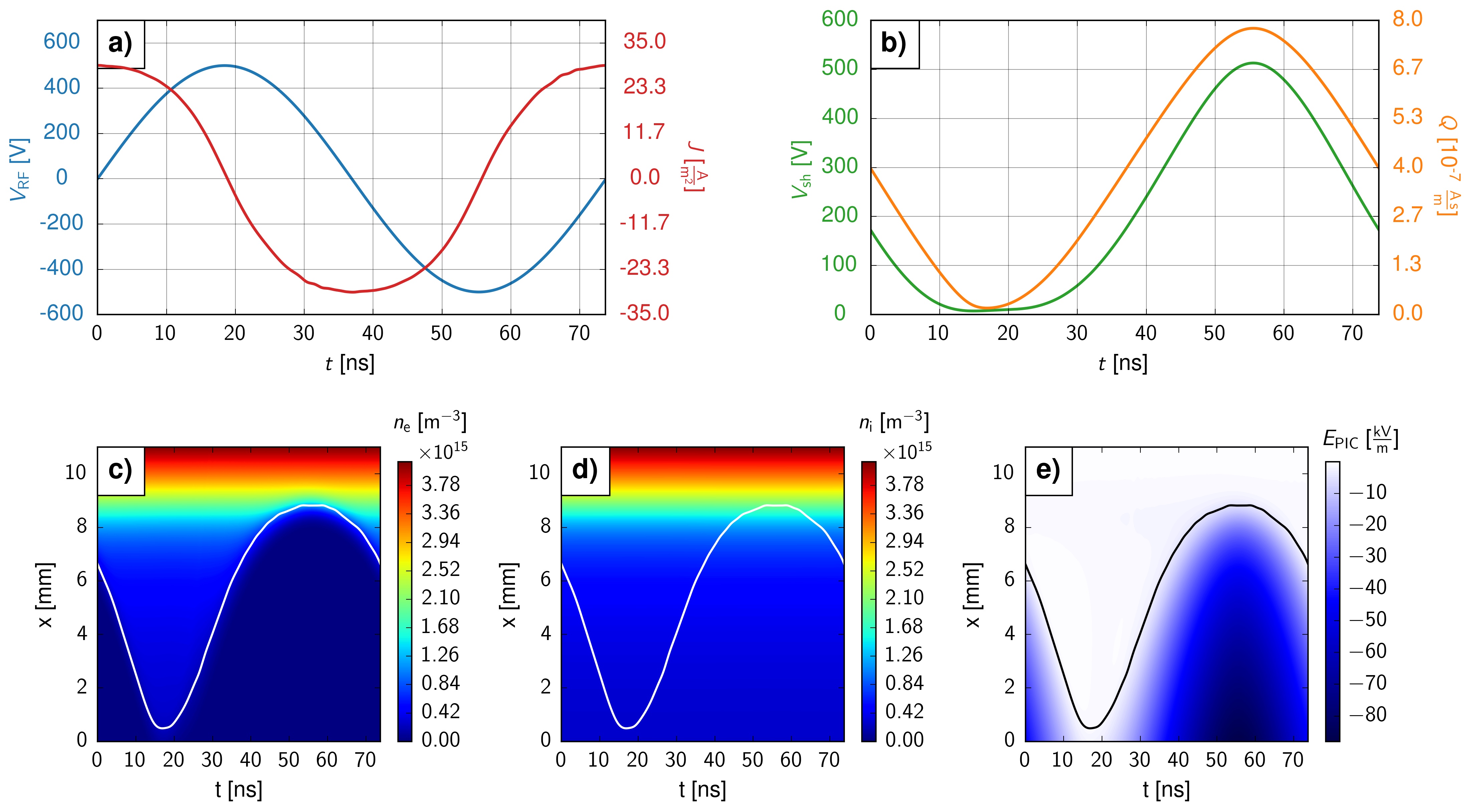}
    \caption{Simulation results for the case 1f.
        a) RF voltage $V_\mathrm{RF}$ (blue) and discharge current $J$ (red).
        \linebreak
          b) sheath voltage $V_\mathrm{sh}$ (green) and sheath charge $Q$ (orange).
        c) to e): spatially and temporally resolved profiles of the electron density $n_\mathrm{e}$, the ion density $n_\mathrm{i}$, and the electric field strength $E_\mathrm{PIC}$.\linebreak The white (in c), d)) and black (in e)) lines mark the position of the formal electron edge $s(t)$. \linebreak
        Conditions: $V_0 = 500\,$V, $\omega_\mathrm{RF} = 2\pi \times 13.56\,$MHz, $L = 50\,$mm, $p = 3\,$Pa.
        }
    \label{current_voltage_1f}
\end{figure}

Case (1f) is driven by $V_\mathrm{RF}(t)= V_0 \sin(\omega_{\mathrm{RF}}t)$, where $V_0 = 500\,\mathrm{V}$ and  $\omega_{\mathrm{RF}}=2\pi\times 13.56\,\mathrm{MHz}$.  
The pressure is $p=3\,\mathrm{Pa}$ at $T_\mathrm{g}=300\,\mathrm{K}$.
The electrode gap of $L = 50\,\mathrm{mm}$ is divided by $600$ grid points,  
the time step is $\Delta t = T_\mathrm{RF}/1000 = 7.4\times 10^{-11}\,\mathrm{s}$. 
Fig.~\ref{current_voltage_1f} a) displays the applied voltage and the resulting current density $J(t)$; the latter is nearly harmonic and has a phase shift to the applied voltage of about $88^\circ$.
 Fig.~\ref{current_voltage_1f} b) shows the sheath charge $Q(t)$ and the sheath voltage $V_\mathrm{sh}(t)$.
The sheath charge $Q(t)$ is obtained by evaluating the right-hand side of equation \eqref{ChargeBalance} for the simulated electric field $E_\mathrm{PIC}$. \linebreak
The sheath voltage results from the spatial integration of the electric field from $0$ to $x_\mathrm{B}$.
\linebreak
Fig.~\ref{current_voltage_1f} c) and d) show the densities of the electrons and ions in the sheath interval $[0,x_\mathrm{B}]$, with
 $x_\mathrm{B} = 11\,\mathrm{mm}$.
The ion density is stationary,  while the electrons are strongly modulated. \linebreak
Fig.~\ref{current_voltage_1f} e) displays the electric field $E(x,t)$.
There is a strong electric field inside the electron depleted zone $x < s(t)$; the field in the ambipolar zone $x > s(t)$ is weak.
The formal electron edge $s(t)$ is marked in the lower panels of Fig.~\ref{current_voltage_1f} in white or black. It is related to the
sheath charge $Q(t)$ by the relation
$q(s(t)) = Q(t)$.

\clearpage

\begin{figure}[h!]
    \centering
    \includegraphics[width=\textwidth]{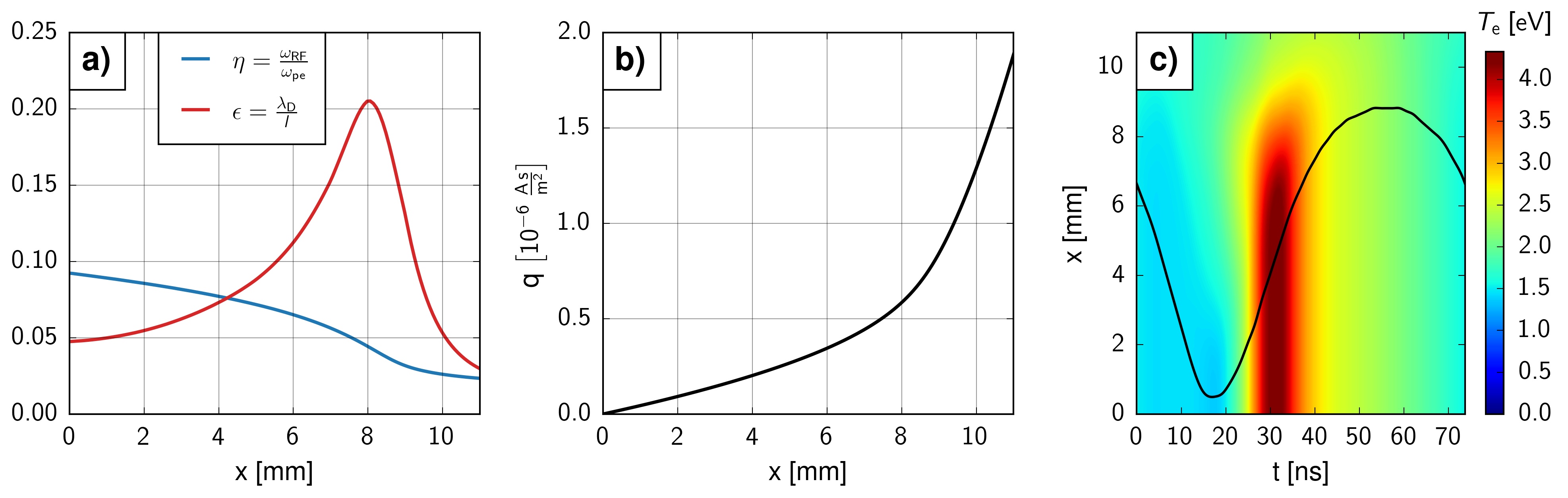}
    \caption{Parameters for the SSM formula calculated from the PIC/MCC simulation of the case 1f.
        a) smallness parameters $\eta$ and $\epsilon$ a function of position $x$.
		b) the charge coordinate function $q(x)$.
		c) the spatially and temporally resolved profile of the regularized electron temperature $T_\mathrm{e}$. 
		The black line marks the position of the formal electron edge $s(t)$.
		Conditions: $V_0 = 500\,$V, $\omega_\mathrm{RF} = 2\pi \times 13.56\,$MHz, $L = 50\,$mm, $p = 3\,$Pa).}
	\label{SSM_paramters_1f}
\end{figure}

Fig.~\ref{SSM_paramters_1f} a) shows the smallness parameters $\epsilon$ and $\eta$ as a function of $x$, evaluated for  time-averaged values of $n_\mathrm{e}$ and $T_\mathrm{e}$.
The frequency ratio $\eta= \omega_\mathrm{RF}/\omega_\mathrm{pe}$ (blue)  is below $0.1$ which indicates quasi-static behavior.
The length scale ratio $\epsilon = \lambda_\mathrm{D}/l$ (red) reaches a maximum of $0.2$.\linebreak
Altogether, the SSM should be a valid approximation. Fig.~\ref{SSM_paramters_1f} b) gives the
function $q(x)$, Fig.~\ref{SSM_paramters_1f} c)
the parallel temperature $T_{\parallel\mathrm{e}}$  
which fills the role of $T_\mathrm{e}$. A regularization procedure was applied which
assigns a finite value in electron-depleted regions.

\begin{figure}[t!]
    \centering
    \includegraphics[width=\textwidth]{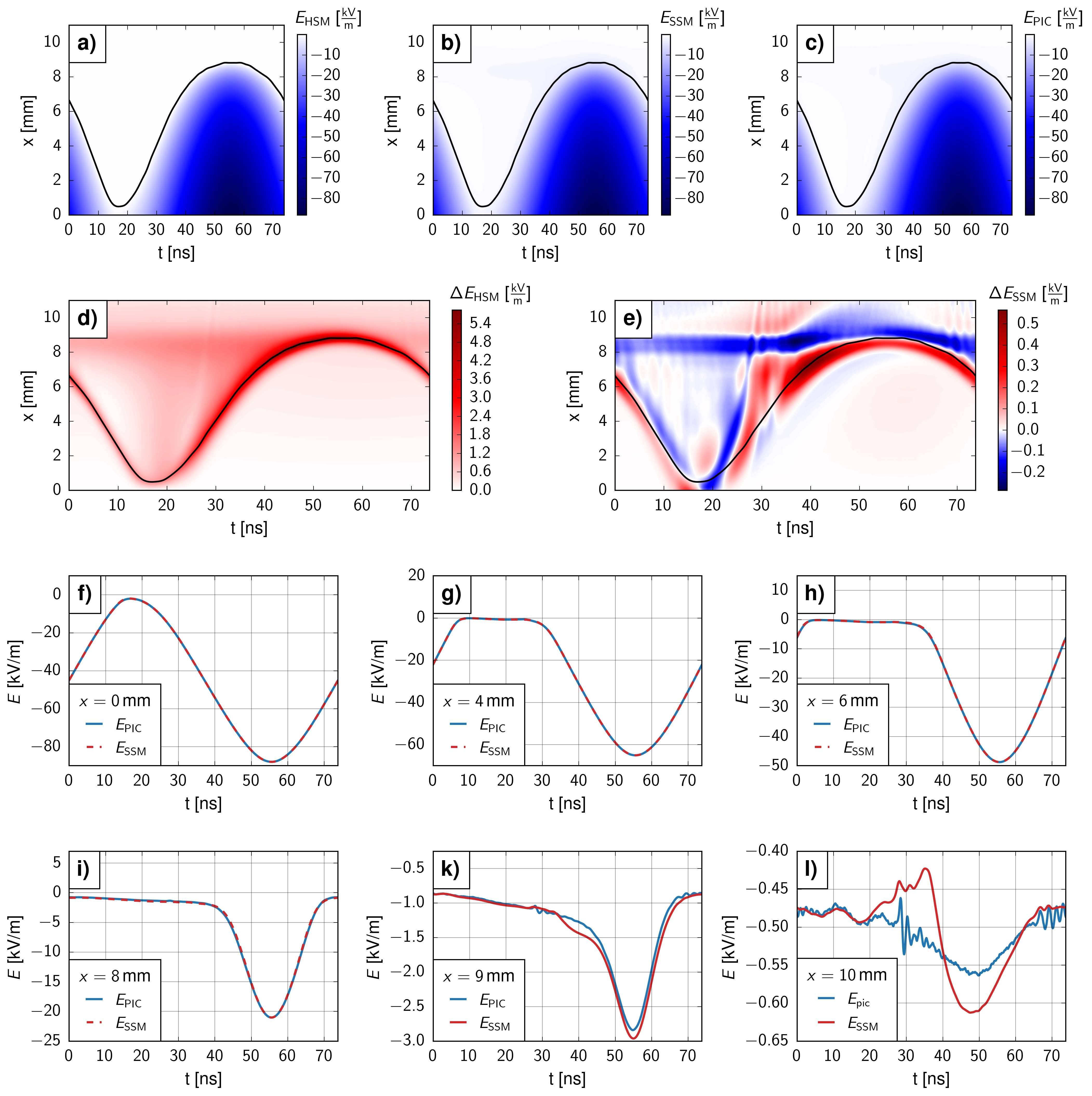}
    \caption{A comparison of the electric field of the case 1f as calculated by the HSM, SSM, and the PIC/MCC simulation.
        a)-c): temporally and spatially resolved profiles of the electric field calculated with a) HSM, b) SSM, c) PIC/MCC.
        d) and e): temporally and spatially resolved profiles of the differences between both models and the simulation data; d) $\Delta E_\mathrm{HSM} = E_\mathrm{HSM} - E_\mathrm{PIC}$, e) $\Delta E_\mathrm{SSM} = E_\mathrm{SSM} - E_\mathrm{PIC}$.
        The black lines mark the position of the formal electron edge $s(t)$.
        f) - l): temporally resolved profiles of the electric field at selected locations within the boundary region displaying the deviation of the SSM and PIC/MCC data in detail. \linebreak
        Conditions: $V_0 = 500\,$V, $\omega_\mathrm{RF} = 2\pi \times 13.56\,$MHz, $L = 50\,$mm, $p = 3\,$Pa).}
    \label{fields_comparison_1f}
\end{figure}

Fig.~\ref{fields_comparison_1f} a) to c) show the temporally and spatially resolved profiles of the electric fields. \linebreak
The HSM, Fig.~\ref{fields_comparison_1f} a) represents only the space charge field,
while the SSM  (Fig.~\ref{fields_comparison_1f} b)) covers the simulation data (Fig.~\ref{fields_comparison_1f} c)) 
in all regions. 
Figure \ref{fields_comparison_1f} d) shows $\Delta E_\mathrm{HSM} = E_\mathrm{HSM} - E_\mathrm{PIC}$.\linebreak 
The PIC/MCC electric field is constantly underestimated.
The SSM has a much lower error, Fig.~\ref{fields_comparison_1f} e) gives $\Delta E_\mathrm{SSM} = E_\mathrm{SSM} - E_\mathrm{PIC}$.
The deviations originate from different sources.\linebreak 
First, the fluid model which underlies the SSM has a limited accuracy.
Then, there is a static error at $x \approx 8\,$mm correlated to the maximum of the length scale ratio $\epsilon$ (Fig.~\ref{SSM_paramters_1f} a)) and some dynamic error around the position of the formal electron edge. 
The structure of the deviations is as expected from previous studies \cite{Brinkmann2015}.
Fig.~\ref{fields_comparison_1f} f) to l) depict the electric field of both the SSM and the PIC/MCC simulation as a function of time for selected positions within the boundary region. 
It shows that the SSM is asymptotically exact in the unipolar zone and very similar to the PIC/MCC field in the ambipolar zone. 

\clearpage

\begin{figure}[t!]
    \centering
    \includegraphics[width=\textwidth]{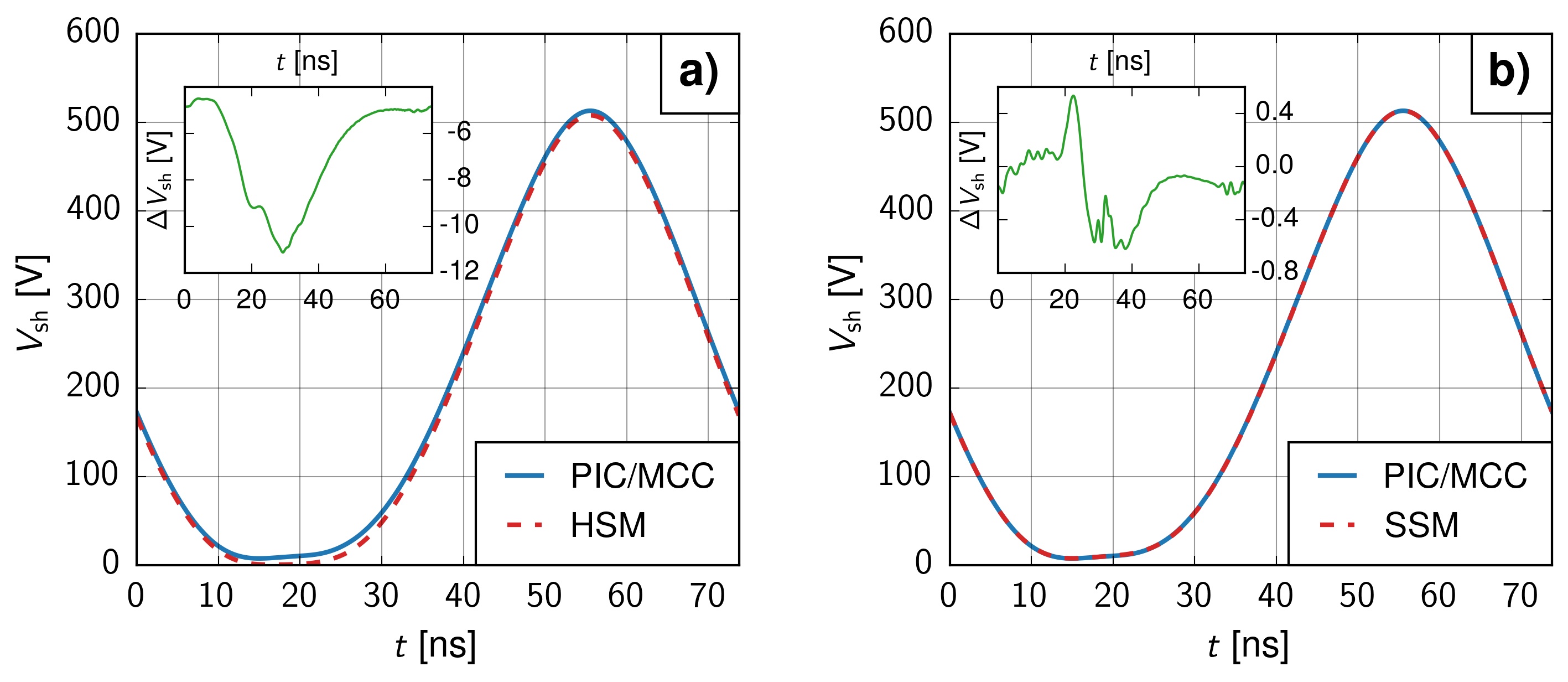}
\caption{Comparison of the model sheath voltages $V_\mathrm{sh}$ to the simulation data for the case (1f).\linebreak
        a): comparison for the HSM. The inset shows the error $\Delta V_\mathrm{sh} = V_\mathrm{sh,HSM} - V_\mathrm{sh,PIC}$.
        b): comparison for the SSM. The inset shows the error $\Delta V_\mathrm{sh} = V_\mathrm{sh,SSM} - V_\mathrm{sh,PIC}$.
        Conditions: $V_0 = 500\,$V, $\omega_\mathrm{RF} = 2\pi \times 13.56\,$MHz, $L = 50\,$mm, $p = 3\,$Pa).    }
    \label{potential_comparison_1f}
\end{figure}

Integrated quantities possess a even smaller error.
Fig.~\ref{potential_comparison_1f} shows a comparison of the predicted sheath voltages a) HSM, b) SSM) to the simulation data.
Already the HSM gives a solid approximation of the sheath voltage $V_\mathrm{sh}$ with an error in the
in the range of a few percent. \linebreak
It misses, however, the physically important contribution of
the ambipolar and Ohmic fields.
The SSM corrects that error, and the  resulting curves  cannot be distinguished (Fig.~\ref{potential_comparison_1f} b)). The inset demonstrates that the difference is indeed very small.
The relative error is less than one percent.

\clearpage

\subsection{Dual frequency case (2f)}
\begin{figure}[h!]
    \centering
    \includegraphics[width=1.0\textwidth]{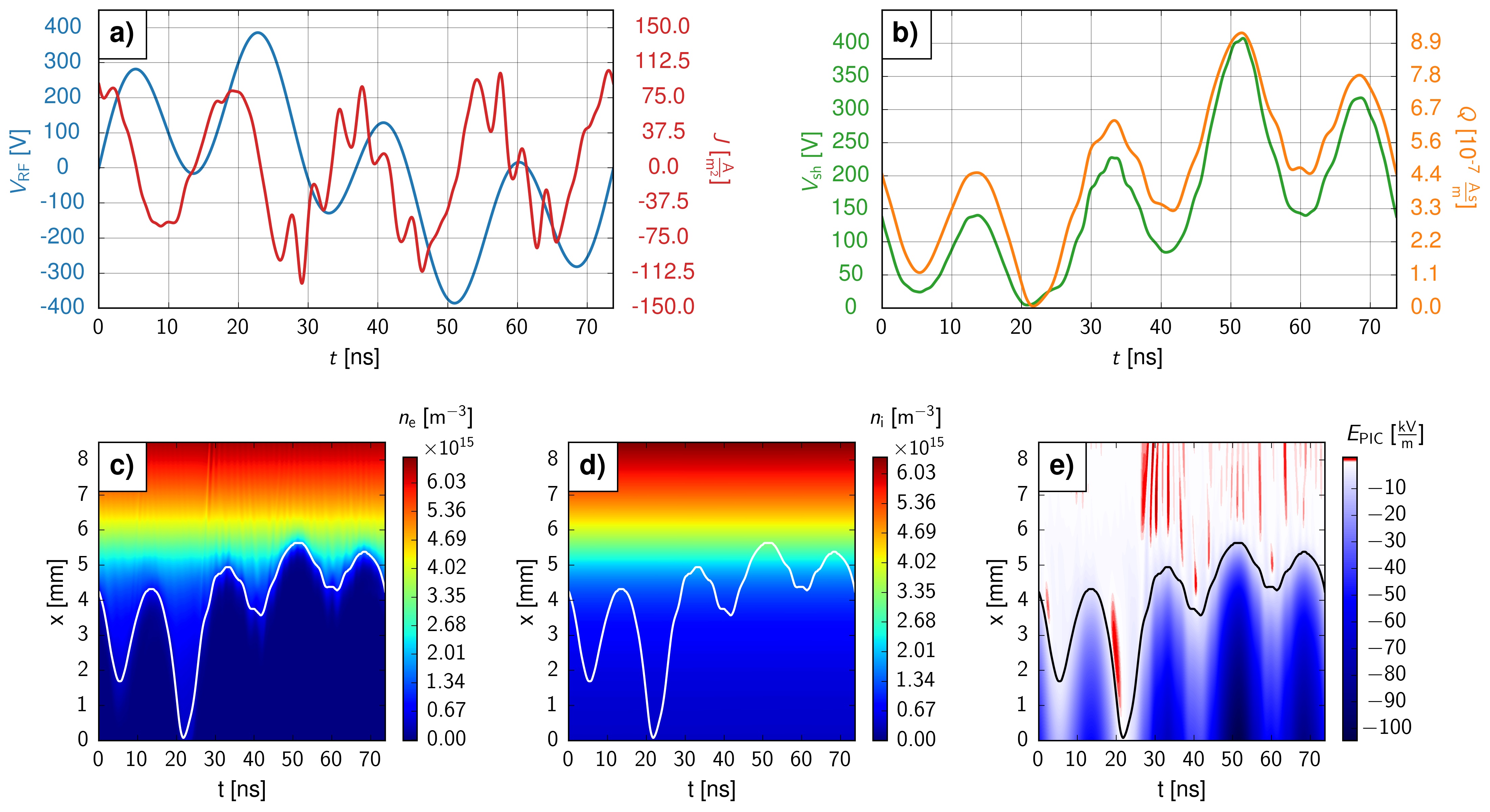}
    \caption{Simulation results for the case 2f.
        a) RF voltage $V_\mathrm{RF}$ (blue) and discharge current $J$ (red). \linebreak
        b) sheath voltage $V_\mathrm{sh}$ (green) and sheath charge $Q$ (orange).
        c) to e): spatially and temporally resolved profiles of c) the electron density $n_\mathrm{e}$, d) the ion density $n_\mathrm{i}$, and e) the electric field $E_\mathrm{PIC}$. \linebreak
        The white (c), d)) and black (e)) lines denote the position of the formal electron edge $s(t)$.\linebreak  Conditions: $V_1 \!=\! V_2 = 200\,$V, $\omega_\mathrm{RF1} = 2\pi \!\times\! 13.56\,$MHz, $\omega_\mathrm{RF2} = 2\pi \!\times\! 54.12\,$MHz, $L = 50\,$mm, $p = 1\,$Pa.\linebreak}
    \label{current_voltage_2f}
\end{figure}

\vspace{-10pt}
Case 2f describes a dual-frequency discharge driven by $V_\mathrm{RF}(t)= 
V_1 \sin(\omega_{\mathrm{RF1}}t)+ V_2 \sin(\omega_{\mathrm{RF2}}t)$,
where $V_1 = V_2 = 200\,\mathrm{V}$, $\omega_{\mathrm{RF1}}=2\pi\times 13.56\,\mathrm{MHz}$, and
$\omega_{\mathrm{RF2}}=2\pi\times 54.12\,\mathrm{MHz}$ (see Fig.~\ref{current_voltage_2f} a)).
\linebreak
The pressure is $p=1\,\mathrm{Pa}$, and the temperature is again fixed to $T_\mathrm{g} = 300\,$K.
The case is a priori chosen to lie at the edge of the validity domain of the SSM. \linebreak
$800$ points discretize the gap of $L = 50\,\mathrm{mm}$; the time step is 
$\Delta t = T_\mathrm{RF1}/2200 = 3.4\times 10^{-11}\,\mathrm{s}$. 
Fig.~\ref{current_voltage_2f} a) shows the discharge
current $J(t)$ which differs strongly from the current of case (1f). \linebreak
In addition to modes in the externally applied frequencies, its shows self-excited  oscillations
at higher frequencies which are the hallmark of the plasma series resonance (PSR) \cite{CzarnetzkiMussenbrockBrinkmann2006}. 
\linebreak
The corresponding sheath charge $Q(t)$ and sheath voltage $V_\mathrm{sh}(t)$ are seen in Fig.~\ref{current_voltage_2f} b).
The~sheath edge is set to $x_\mathrm{B} = 8\,\mathrm{mm}$. Figs.~\ref{current_voltage_2f} c), d) show the densities in the interval $[0,x_\mathrm{B}]$. \linebreak
Again, the ion density (Fig.~\ref{current_voltage_2f} d)) is not RF modulated. The electron density (Fig.~\ref{current_voltage_2f} c)) is not only strongly modulated, but also  shows fast electrostatic plasma oscillations \cite{TonksLangmuir1929}.
\linebreak
Fig.~\ref{current_voltage_2f} e) shows the electric field $E(x,t)$ of case 2f.
The electron edge $s(t)$ is depicted in the lower panels as a white (c), d)) or black curve(e)), respectively.
In contrast to case 1f (c.f., Fig.~\ref{current_voltage_1f} e)), the electric field shows strong oscillations in the ambipolar zone in front of the electron edge ($x > s(t)$).
Similar structures show in the electron density (Fig.~\ref{current_voltage_2f} c)).

\clearpage

\begin{figure}[h!]
    \centering
    \includegraphics[width=\textwidth]{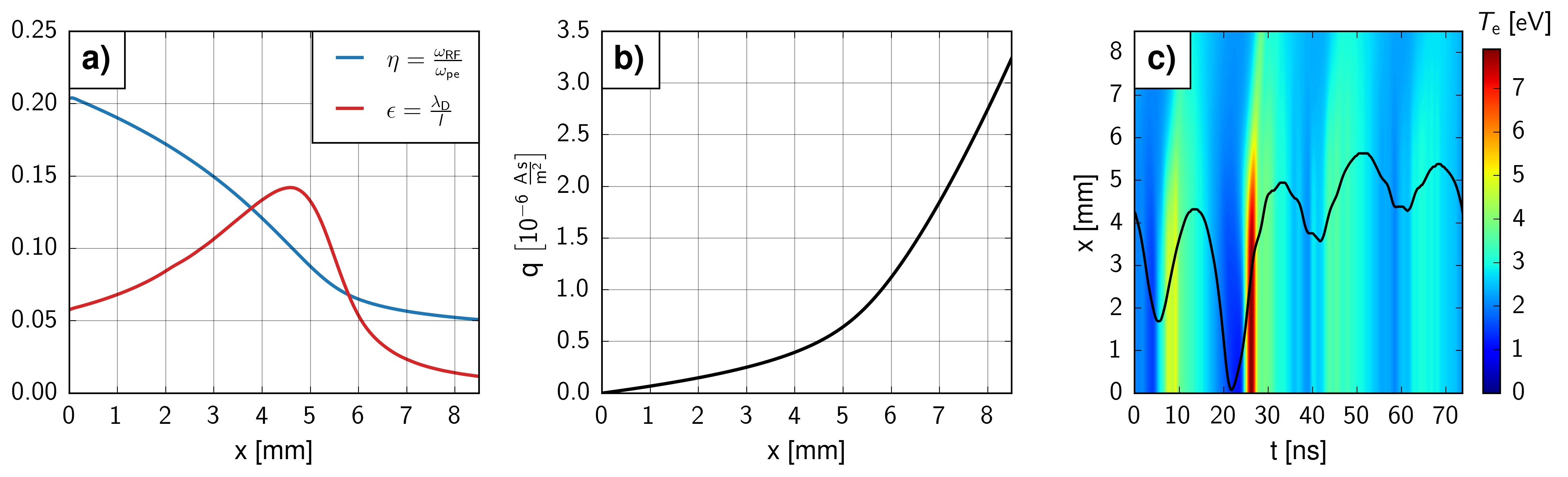}
    \caption{Parameters for the SSM formula calculated from the PIC/MCC simulation of the case 2f.
        a) smallness parameters $\eta$ and $\epsilon$ a function of position $x$.
		b) the charge coordinate function $q(x)$. \linebreak
		c) the spatially and temporally resolved profile of the regularized electron temperature $T_\mathrm{e}$.\linebreak
		The black line marks the position of the formal electron edge $s(t)$.
		Conditions: $V_1 = V_2 = 200\,$V, $\omega_\mathrm{RF1} = 2\pi \times 13.56\,$MHz, $\omega_\mathrm{RF2} = 2\pi \times 54.12\,$MHz, $L = 50\,$mm, $p = 1\,$Pa.}
	\label{SSM_parameters_2f}
\end{figure}

Fig.~\ref{SSM_parameters_2f} a) depicts the spatial distribution of the smallness parameters
$\epsilon$ and $\eta$. The assumption of quasistatic behavior is particularly questionable, 
as the frequency ratio $\eta$ is above 0.2 and thus fairly large.
The length scale ratio $\epsilon$ behaves comparable to the case 1f and just displays the expected maximum at $x \approx 4.5\,$mm.
Again, most quantities which enter the SSM formula were already displayed. The remaining quantities are the function $q(x)$,
Fig.~\ref{SSM_parameters_2f} b), and the regularized parallel temperature $T_{\parallel\mathrm{e}}$ which assumes the role of $T_\mathrm{e}$,
Fig.~\ref{SSM_parameters_2f} c).

\pagebreak

\begin{figure}[h!]
    \centering
    \includegraphics[width=0.97\textwidth]{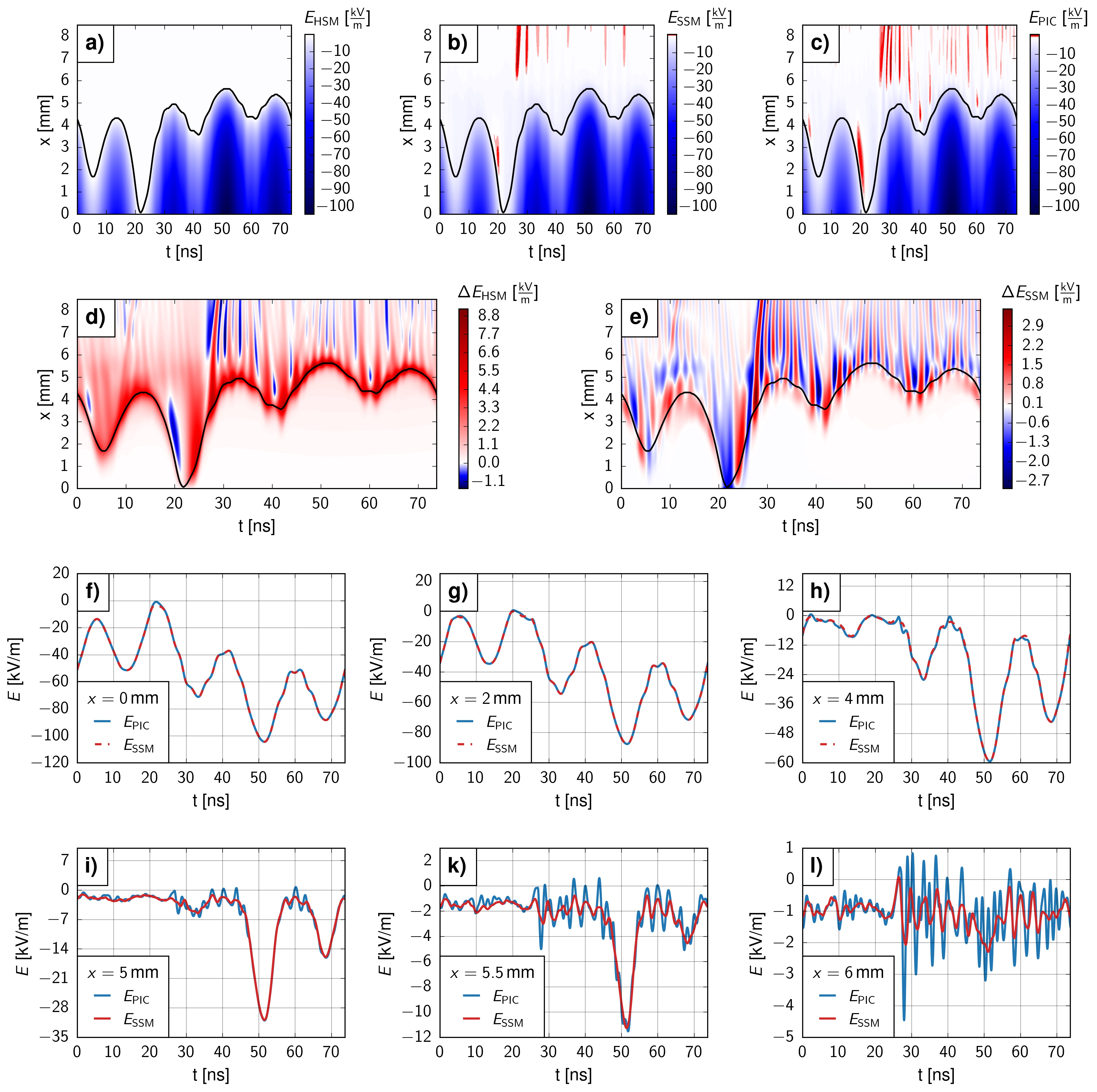}
\caption{Electric fields of the case (2f) as calculated by the HSM, the SSM, and the PIC/MCC.
Figs.~a) - c) show the temporally and spatially resolved profiles of the electric field.
        d) and e): temporally and spatially resolved profiles of the differences between both models and the simulation data; d) $\Delta E_\mathrm{HSM} = E_\mathrm{HSM} - E_\mathrm{PIC}$, e) $\Delta E_\mathrm{SSM} = E_\mathrm{SSM} - E_\mathrm{PIC}$.
        The black lines mark the position of the formal electron edge $s(t)$.
        f) - l): temporally resolved profiles of the electric field at selected locations within the boundary region displaying the deviation of the SSM and PIC/MCC data in detail. \linebreak
        Conditions: $V_1 \!=\! V_2 = 200\,$V, $\omega_\mathrm{RF1} = 2\pi \!\times\! 13.56\,$MHz, $\omega_\mathrm{RF2} = 2\pi \!\times\! 54.12\,$MHz, $L = 50\,$mm, $p = 1\,$Pa.\linebreak}
    \label{fields_comparison_2f}
\end{figure}

\pagebreak

Fig.~\ref{fields_comparison_2f} a) to c) depict temporally and spatially resolved profiles of the electric field $E(x,t)$.
It becomes evident  that both the HSM (Fig.~\ref{fields_comparison_2f} a)) and the SSM (Fig.~\ref{fields_comparison_2f} b)) show deviations from the simulation data (Fig.~\ref{fields_comparison_2f} c)).
Temporally and spatially resolved profiles of these deviations are provided in figure \ref{fields_comparison_2f} d) (for the HSM $\Delta E_\mathrm{HSM} = E_\mathrm{HSM} - E_\mathrm{PIC}$) and e) (for the SSM $\Delta E_\mathrm{SSM} = E_\mathrm{SSM} - E_\mathrm{PIC}$).
The HSM, again, misses the Ohmic and the ambipolar field, and exhibits the largest error around the sheath edge.
Additionally, strong oscillations are visible for $x>5\,$mm.
They stem from the PIC/MCC simulation exhibiting superimposed oscillations which the HSM is unable to resolve. 

As~expected, the SSM is still exact in the unipolar region, see Fig.~\ref{fields_comparison_2f} e).
However, in the ambipolar regions there are considerable deviations as well and the overall error is larger than for the case 1f.
The static error connected to the maximum in the length scale ratio $\epsilon$ and the curvature of the ion density $n_\mathrm{i}$ is probably overshadowed by the large dynamic errors. \linebreak
The oscillations found in the simulated field on the timescale of the plasma frequency $\omega_\mathrm{pe}$
cannot be resolved by the SSM which is based on the quasi-static 
assumption.

Figure \ref{fields_comparison_2f} f) to l) present temporal profiles of the electric field calculated by both the SSM and the PIC/MCC simulation for case 2f.
Panels f) to h) prove the exactness of the SSM within the unipolar region.
The panels i) to l) display the inability of the SSM to follow the oscillations mentioned above.
Resulting from this inability, the SSM assumes some kind of mean value whenever unable to keep track of high-frequent oscillations resulting in considerably larger errors. 

\clearpage

\begin{figure}[h!]
    \centering
    \includegraphics[width=\textwidth]{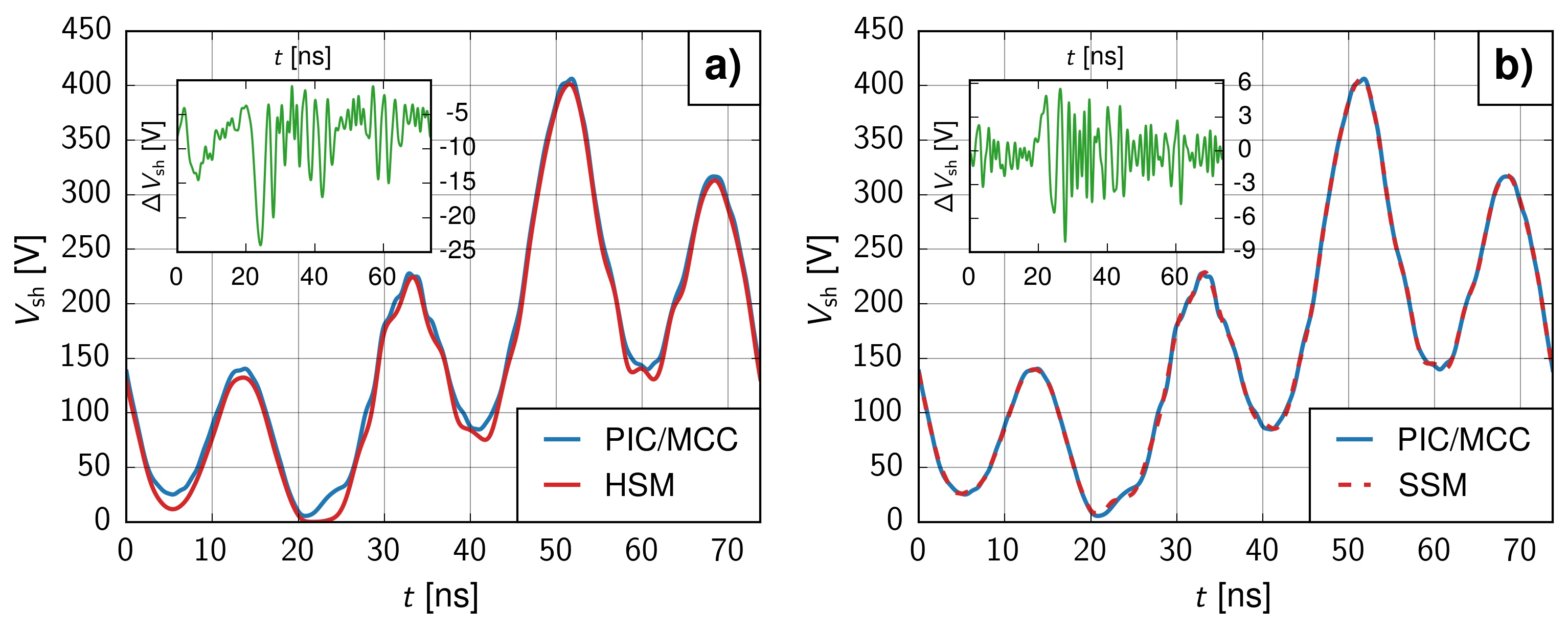}
\caption{Comparison of the model sheath voltages $V_\mathrm{sh}$ with the PIC/MCC simulation for case~(2f).
        a) HSM, inset giving $\Delta V_\mathrm{sh} = V_\mathrm{sh,HSM} - V_\mathrm{sh,PIC}$.
        b) SSM, inset giving $\Delta V_\mathrm{sh} = V_\mathrm{sh,SSM} - V_\mathrm{sh,PIC}$.
        Conditions: $V_1 \!=\! V_2 = 200\,$V, $\omega_\mathrm{RF1} = 2\pi \!\times\! 13.56\,$MHz, $\omega_\mathrm{RF2} = 2\pi \!\times\! 54.12\,$MHz, $L = 50\,$mm, $p = 1\,$Pa.}
    \label{potential_comparison_2f}
\end{figure}

Figure \ref{potential_comparison_2f} shows a comparison of the sheath voltages predicted by the HSM (Fig.~\ref{potential_comparison_2f} a)) and the SSM (Fig.~\ref{potential_comparison_2f} b) to the PIC/MCC simulation.
For the HSM, the differences are evident and the inset shows that the sheath voltage is severely underestimated for this case.
The largest error occurs between $20$ to $25\,$ns and is close to 100 percent.
In contrast, global quantities such as the sheath voltage are still quite reasonably given by the SSM formula, even though the underlying assumptions are only marginally met.
Figure \ref{potential_comparison_2f} b) shows nearly identical curves and the inset supports this impression. 
The overall error in the sheath voltage does not exceed a few percent.

\pagebreak

\section{Conclusion and outlook}

The smooth step model (SSM) is an approximate solution of the electron equations of motion and Poissons's equation in the RF regime
$\omega_\mathrm{pi}\!\ll\!\omega_\mathrm{RF}\!\ll\! \omega_\mathrm{pe}$.  
The subject of this manuscript is the validation of the SSM, using simulations with a particle-in-cell/Monte Carlo collisions (PIC/MCC) code as reference.
Comparisons are also made with the hard step model (HSM). 
Two exemplary  discharge configurations were chosen. 
Both employ the same capacitively coupled parallel plate reactor with a gap width
of $L = 50\,\mathrm{mm}$. The single frequency case (1f)\linebreak  operates with $\omega_{\mathrm{RF}}=2\pi\times 13.56\,\mathrm{MHz}$
at a pressure of $p = 3\,\mathrm{Pa}$; the dual frequency case (2f) 
is run with $\omega_{\mathrm{RF1}}=2\pi\times 13.56\,\mathrm{MHz}$,
$\omega_{\mathrm{RF2}}=2\pi\times 54.12\,\mathrm{MHz}$ at $p = 1\,\mathrm{Pa}$. 

The SSM performs as expected: In the (1f) case, where the prior assumptions 
are well met, the deviations from the PIC/MCC standard are small. 
The electric field is captured well,\linebreak both phase-resolved and phase-averaged.
In the unipolar zone, it is asymptotically~exact, a feature shared with the HSM.
In the ambipolar region, it shows an error which can be attributed to the underlying fluid model:
The description of the collisional momentum loss by a friction ansatz
is limited, especially at low-pressure. Nonetheless, the SSM outperforms the HSM
which neglects Ohmic and ambipolar fields altogether. The error caused by the approximation technique itself
(i.e., the use of a truncated asymptotic series in the smallness parameters
$\eta = \omega_\mathrm{RF}/\omega_\mathrm{pe}$
and $\epsilon = \lambda_\mathrm{D}/l$) is negligible in comparison.
The sheath voltage as a spatially integrated quantity has an error of less than 1\,\%.

The situation is different in the case (2f). Here, the smallness parameter $\eta$ reaches 20\%, \linebreak
which indicates that the assumption of quasi-static behavior is only marginally fulfilled. \linebreak
And indeed,  high-frequent oscillations on the scale of the plasma frequency are observed in the ambipolar zone. 
The SSM is, by nature, not able to follow these fast oscillations and results in a temporally  averaged field. 
(Again, the SSM performs better than the HSM which predicts a vanishing field.) 
The overall sheath voltage is still captured well.

All in all, we state that the validation of the SSM was successful. The approximation is valid \linebreak in the
regime that it is designed for, namely in the RF regime where $\omega_\mathrm{RF}\ll \omega_\mathrm{pe}$
and $\lambda_\mathrm{D}\ll l$.
(Only the first condition is critical; the second is self-adjusting as the gradient length of the ion density  $l$
itself scales with $\lambda_\mathrm{D}$.) The SSM eclipses the HSM.  
The approximation is thus ready to be used for the modeling of RF modulated gas discharges, for example for theories of stochastic heating and for the definition of 
effective boundary sheath models 
\cite{Brinkmann2016,SamirWilczekKlichMussenbrockBrinkmann2021}.

\pagebreak

\section*{Acknowledgement}

Funded by the Deutsche Forschungsgemeinschaft (DFG, German Research Foundation) – Project-ID 327886311 (CRC 1316).

\section*{ORCID IDs:}
\noindent M.~Klich: \href{https://orcid.org/0000-0002-3913-1783}{https://orcid.org/0000-0002-3913-1783}

\noindent S.~Wilczek: \href{https://orcid.org/0000-0003-0583-4613}{https://orcid.org/0000-0003-0583-4613}

\noindent T.~Mussenbrock: \href{http://orcid.org/0000-0001-6445-4990}{http://orcid.org/0000-0001-6445-4990}

\noindent R.P.~Brinkmann: \href{https://orcid.org/0000-0002-2581-9894}{https://orcid.org/0000-0002-2581-9894}

\end{document}